\documentclass[preprint,authoryear,12pt]{elsarticle}

\usepackage{epsfig}

\usepackage{amssymb}

\journal{Journal of Atmospheric and Solar-Terrestrial Physics}

\begin{document}
\begin{frontmatter}


\author{Jan-Erik Solheim\corref{cor1}}
\ead{janesol@online.no}
\cortext[cor1]{Corresponding Author}
\address{Department of Physics and Technology, University of Troms\o, N-9037, Troms\o, Norway}
\author{Kjell Stordahl}
\address{Telenor Norway, Fornebu, Norway}
\ead{Kjell.Stordahl@telenor.com}
\author{Ole Humlum}
\address{Department of Geosciences, University of Oslo, Norway}
\address{Department of Geology, University Centre in Svalbard (UNIS), Svalbard}
\ead{Ole.Humlum@geo.uio.no}

\title{The long sunspot cycle 23 predicts a significant temperature decrease in cycle 24}


\author{}
\address{}

\begin{abstract}
Relations between the length of a sunspot cycle and the average temperature in the same and the next cycle are calculated for a number of meteorological stations in Norway and in the North Atlantic region. 
No significant trend is found between the length of a cycle and the average temperature in the same cycle, but a significant negative trend is found between the length of a cycle and the temperature in the next cycle.
This provides a tool to predict an average temperature decrease of at least 1.0 $ ^{\circ}$C from solar cycle 23 to 24 for the stations and areas analyzed. 
 We find for the Norwegian local stations investigated that 25--56\% of the temperature increase the last 150 years may be attributed to the Sun. For 3 North Atlantic stations we get 63--72\% solar contribution. This points to the Atlantic currents as reinforcing a solar signal.

\end{abstract}

\begin{keyword}
Solar cycle lengths, Climate change, Forecasts, North Atlantic climate response

\end{keyword}

\end{frontmatter}


\section{Introduction}
\label{Intro}

The question of a possible relation between solar activity and the Earth's climate has received considerable attention during the last 200 years. Periods with many sunspots and faculae correspond with periods with higher irradiance in the visual spectrum and even stronger response in the ultraviolet, which acts on the ozone level. 
It is also proposed that galactic cosmic rays can act as cloud condensation nuclei, which may link variations in the cloud coverage to solar activity, since more cosmic rays penetrate the Earth's magnetic field when the solar activity is low. A review of possible connections between the Sun and the Earth's climate is given by \citet{Gr10}. 
 
Based on strong correlation between the production rate of the cosmogenic nucleids $^{14}$C and $^{10}$Be and proxies for sea ice drift, \citet{Bo01} concluded that extremely weak pertubations in the Sun's energy output on decadal to millennial time scales generate a strong climate response in the North Atlantic deep water (NADW). This affects the global thermohaline circulation and the global climate. The possible sun-ocean-climate connection may be detectable in temperature series from the North Atlantic region. 
  Since the ocean with its large heat capacity can store and transport huge amounts of heat, a time lag between solar activity and air temperature increase is expected. 
  An observed time lag gives us an opportunity for forecasting, which is the rationale for the present investigation.

Comparing sunspot numbers with the Northern Hemisphere land temperature anomaly, \citet{FCL91} noticed a similar behavior of temperature and sunspot numbers from 1861 to 1990, but it seemed that the sunspot number $ R $ appeared to lag the temperature anomaly. 
They found a much better correlation between the solar cycle length (SCL) and the temperature anomaly. 
In their study they used a smoothed mean value for the SCL with 5 solar cycles weighted 1-2-2-2-1. They correlated the temperature during the central sunspot cycle of the filter with this smoothed weighted mean value for SCL. The reason for choosing this type of filter was that it has traditionally been used to describe long time trends in solar activity. However, it is surprising that the temperature was not smoothed the same way.
In a follow up paper \citet{RT01} concluded that the right cause-and-effect ordering, in the sense of Granger causality, is present between the smoothed SCL and the cycle mean temperature anomaly for the Northern Hemisphere land air temperature in the twentieth century at the 99\% significance level. This suggests that there may exist a physical mechanism linking solar activity to climate variations. 

\begin{figure}
\epsfysize=40mm 
\epsfbox{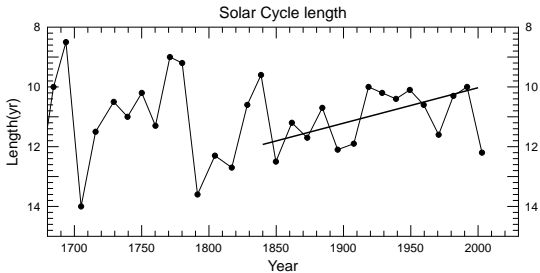}
\caption{Length of solar cycles (inverted) 1700-2009.The last point refers to SC23 which is 12.2 years long. The gradual decrease in solar cycle length 1850-2000 is indicated with a straight line.}
\end{figure}

The length of a solar cycle is determined as the time from the appearance of the first spot in a cycle at high solar latitude, to the disappearance of the last spot in the same cycle near the solar equator. However, before the last spot in a cycle disappears, the first spot in the next cycle appears at high latitude, and there is normally a two years overlap. 
The time of the minimum is defined as the central time of overlap between the two cycles \citep{Wa39}, and the length of a cycle can be measured between successive minima or maxima.

A recent description of how the time of minimum is calculated is given by \citet{ng11}:
"When observations permit, a date selected as either a cycle minimum or maximum is based in part on an average of the times extremes are reached in the monthly mean sunspot number, in the smoothed monthly mean sunspot number, and in the monthly mean number of spot groups alone.
Two more measures are used at time of sunspot minimum: the number of spotless days and the frequency of occurrence of old and new cycle spot groups."
	
It was for a long time thought that the appearance of a solar cycle was a random event, which means for each cycle length and amplitude were independent of the previous.  However, \citet{Di78} showed that an internal chronometer has to exist inside the Sun, which after a number of short cycles, reset the cycle length so the average length of 11.2 years is kept.	
\citet{RR09} analyzed the length of cycles 1610-2000 using median trace analyses of the cycle lengths and power spectrum analyses of the $ O-C $ residuals of the dates of sunspot maxima and minima. They identified a period of $ 188\pm 38 $ yr. They also found a correspondence between long cycles and minima of number of spots. 
Their study suggests that the length of sunspot cycles should increase gradually over the next $ \sim 75$ yr. accompanied by a gradual decrease in the number of sunspots.

An autocorrelation study by \citet{SK02} showed that the length of a solar cycle is a good predictor for the maximum sunspot number in the next cycle, in the sense that short cycles predict high $ R_{max} $ and long cycles predict small $ R_{max} $. They explain this with the solar dynamo having a memory of the previous cycle's length.

Assuming a relation between the sunspot number and global temperature, the secular periodic change of SCL may then correlate with the global temperature, and as long as we are on the ascending (or descending) branches of the 188 yr period, we may predict a warmer (or cooler) climate.

It was also demonstrated \citep{FCL92,HS93,LFC95} that the correlation between SCL and climate probably has been in operation for centuries. A statistical study of 69 tree rings sets, covering more than 594 years, and SCL demonstrated that wider tree-rings (better growth conditions) were associated with shorter sunspot cycles \citep{ZB98}.

The relation between the smoothed SCL and temperature worked well as long as SCL decreased as shown in figure 1. But when the short cycle SC22 was finished \citet{TL00} reported a developing inconsistency. 
In order to explain the high temperatures at the turn of the millennium, the not yet finished SC23 had to be shorter than 8 years, which was very unlikely, since there had never been observed two such short cycles in a row (see fig. 1). 
They concluded that the type of solar forcing described with this SCL model had ceased to dominate the temperature change. Since the final length of SC23 became 12.2 yrs, the discrepancy became even bigger. 

A more rigorous analysis was done by \citet{Th09}, this time with a filter 1-2-1 also used for temperature. In particular he investigated if the residuals from the relation between Northern Hemisphere (NH) land ($t_{121} $) and SCL$ _{121} $ were independent and identically distributed. 
He also investigated possible lags between SCL$ _{121} $ and $t_{121}$, and found that time lags of 6 or 12 years were necessary to obtain zero serial correlation in residuals in the data set used by \citet{FCL91}. For updated NH-land series from the HadCRU Centre, serial correlation in residuals could not be removed \citep{Th09}.

In addition to the relation between solar cycle length and the amplitude of the next $ R_{max} $, it is reasonable to expect a time lag for the locations investigated, since heat from the Sun, amplified by various mechanisms, is stored in the ocean mainly near the Equator, and transported into the North Atlantic by the Gulf Stream to the coasts of Northern Europe. 
An example of time lags along the Norwegian coast is an advective delay between the Faroe-Shetland Channel and the Barents Sea of about 2 years determined from sea temperature measurements \citep{YT08}.

Formation of NADW represents transfer of upper level water to large depths.
 The  water is transported and spread throughout the Atlantic and exported to the Indian and Pacific oceans before updwelling in Antarcic waters. 
The return flow of warm water from the Pacific through the Indian ocean and the Caribbean to the North Atlantic, a distance of 40\,000 km, takes from 13 to 130 years \citep{Go86}. 
There appears to be solar "fingerprints" that can be detected in climate time series in other parts of the world with each series having a unique time lag between the solar signal and the hydro-climatic response. 
\citet{Pe07} reports that a solar signal composed of geomagnetic  $ aa $-index and total solar irradiance (TSI) is detected with various lags from zero years (Indian Ocean) to 34 years (Mississippi river flow) and 70 years (Labrador Sea ice).
\citet{Me09} have shown that two mechanisms: the top-down stratospheric response of ozone to fluctuations of shortwave solar forcing and the bottom-up coupled ocean-atmospheric surface response, acting together, can amplify a solar cyclical pulse with a factor 4 or more.
Since our stations are located near or in the North Atlantic, solar signals in climatic time series may arrive with delays of decades. If we can detect a solar signal and measure the delay for individual regions, we may have a method for future climate predictions. 

Recognizing that averaged temperature series from different meteorological stations of variable quality and changing locations may contain errors, and partially unknown phenomena derived from the averaging procedure, \citet{Bu94} proposed instead to use long series of high quality single stations. 
This might improve the correlation between SCL and temperature.

He showed that this was the case when using temperature series obtained at the Armagh Observatory in Northern Ireland 1844-1990. 
Since the Armagh series correlated strongly with the NH temperature, he concluded that this indicates that solar activity, or something closely related to it, has had a dominant influence on the temperature of the lower atmosphere in the Northern Hemisphere over the past 149 years \citep{Bu94}. 
This investigation was later expanded to the period 1795-1995 by \citet{BJ94}, who found a relation $ t_{ARM} = 14.42 - 0.5L_{SC}$, where $ L_{SC} $ is the smoothed length of the solar cycle determined by the 1-2-2-2-1 filter. 
They concluded that the good fit over nearly 200 years indicated that solar activity had been the dominant factor for nearly two centuries.  

 \citet{BJ96}, studying the same dataset, noticed that there seemed to be about one solar cycle delay between the shortest cycle lengths and the temperature peaks. 
 They therefore compared the smoothed value of SCL with the temperature 11 years later, whereby the correlations improved. 
 They also compared the correlations with the raw, (unsmoothed) times of minima and maxima, filter 1-2-1, and the 1-2-2-2-1 filter, and found that relations were approximately the same, but the correlations improved significantly with the length of the filter. 
They interpreted this as the link between the solar dynamo and the mean air temperature at Armagh is a gradual process that becomes more evident when SCL is smoothed over several sunspot cycles. They also found that temperatures shifted 11 years back in time, correlated better with SCL measured between minima than between maxima.

In figure 1, the length of sunspot cycles between successive minima is plotted inverted. 
Peaks in this curve preceded peaks in the Armagh temperature curve, and bumps preceded colder periods.
A straight line with slope
$ \beta = -(0.012\pm0.004 ) $yr$ ^{-1} $ represents the average change in sunspot cycle length in the period 1843-1996.

Another study by \citet{Wi98} presented extremely good correlations between the temperature series at Armagh and the NH-temperatures (r=0.88) and global temperatures (r=0.83), and explained this with the influence of the ocean. He also found an extremely good negative correlation between the length of the sunspot Hale cycle of about 23 years and the Armagh temperature average. In addition to this he also found a good correlation between even and odd sunspot cycles and predicted t=9.24$ \pm 0.47^{\circ}$C (90\% confidence interval) for SC23. The observed value became 10.1$^{\circ}$C - somewhat higher than predicted.

\citet{AR08} was the first to realize that the length of the previous sunspot cycle (PSCL) has a predictive power for the temperature in the next sunspot cycle, if the raw (unsmoothed) value for the SCL is used. 
Based on the estimated  length of SC23 then being 12.6 years, considerably longer than SC22 of 9.6 years\footnote{The recent version of SCL numbers obtained from \citet{ng11}, used in this work, shows SCL22 = 10.0 and SCL23 = 12.2 years between minima.}, he predicted cooling during the coming SC24 for certain locations.

He demonstrated this based on a long series from de Bilt in the Netherlands 1705-2000 which showed a decrease of 0.6$ ^{\circ}$C per year PSCL and for Hanover NH with a slope of  -0.73$ ^{\circ}$C yr$ ^{-1} $. Other long series he investigated was Portland, ME (slope: -0.70$ ^{\circ}$C yr$ ^{-1} $), Providence, RI (slope: -0.62 $ ^{\circ}$C yr$ ^{-1} $), and Archangel, RU with a slope -0.6 $ ^{\circ}$C yr$ ^{-1} $ PSCL.
In his analysis he used both the times between maxima and between minima in the same relations \citep{AR10}.

 In the following we will compare the relations between raw (unsmoothed) SCL values and temperatures in the same or in the next sunspot cycles. The use of unsmoothed SLC values may give a better response to the large change in SCL that took place from SC22 to SC23. 
We will also investigate the relations using different time lags.
Our main goal is to investigate the use of PSCL as a tool for temperature predictions at certain locations. 
In addition we will use the relation between PSCL and temperature to estimate how much of the temperature variations in the series investigated, may be attributed to solar activity. We will also use the PSCL-relations to predict the future temperature changes from SC23 to SC24 - a prediction that may be falsified in less than 10 years. In section 2 we present data sets and methods. In section 3 results, and then discussion and conclusions in sections 4 and 5.
   
\section{Data sets and method}
\begin{figure}[htb]
\epsfysize=60mm 
\epsfbox{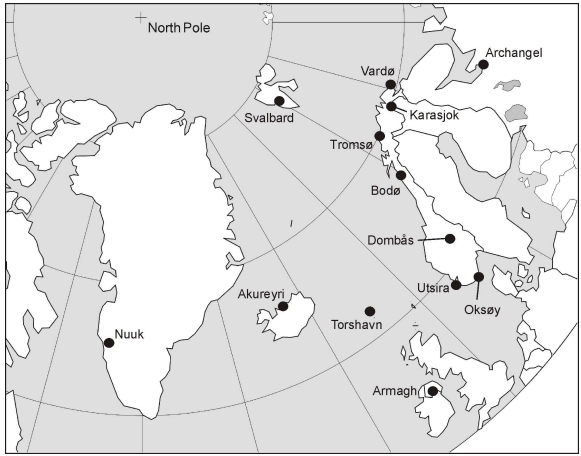}
\caption{The location of meteorological stations with temperature series analyzed.}
\end{figure}
Mean annual air temperatures during sunspot cycles, as far as possible back to 1856, which was the start of SC10, are calculated for a selection of Norwegian weather stations, which have long records. 
Both costal and inland stations are represented. 
To avoid urban heat effects, only temperature series from places with small populations or stations at lighthouses are analyzed.  
These are then correlated with the length of the same (SCL) and the previous (PSCL) sunspot cycles. 
Based on relations determined between PSCL and temperature, temperature forecast for solar cycle 24 are calculated with 95\% confidence interval  when possible.
To investigate the validity of the relations found, also North Atlantic temperature series from Armagh, Archangel, The Faroe Islands, Iceland, Svalbard and Greenland are analyzed. The location of the stations used is shown in figure 2.

We have also determined SCL-- and PSCL--relations for composite temperature averages for Norway and we have constructed a European mean temperature anomaly based on 60 stations (Europe60), listed and shown on a map (figure A.20) in the appendix. This composite series has an increasing number of stations included with time, as shown in figure A.21.
These stations were selected to 1) obtain long data series, 2) obtain good geographical coverage, and 3) to minimize potential effects from urban heat islands.
In addition we have investigated the same relations for HadCRUT3N temperature anomalies.

As an estimate of a possible solar effect related to  PSCL, the coefficient of determination $ r^{2} $ is calculated, where $ r $is the correlation coefficient. 
 In addition we have investigated  present autocorrelations in the residuals when the relations between PSCL and temperature is employed, to examine the level of  presence of other possible regressors. 
 Finally we have also tested our forecasts by removing the last observation and made a forecast based on the remaining data points. 
 
\begin{figure}[htb]
\epsfysize=70mm 
\epsfbox{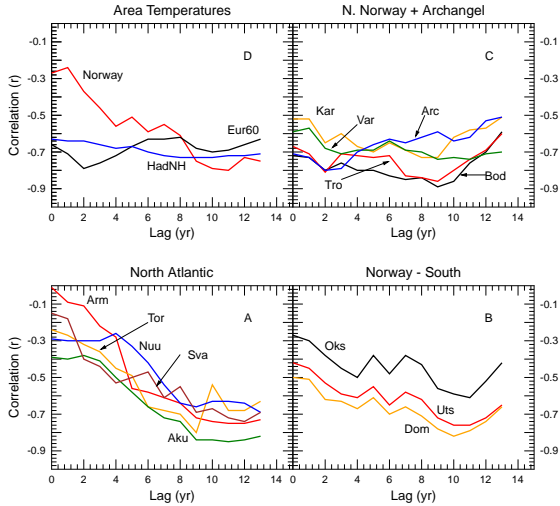}
\caption{Correlations between SCL and 11 years average temperatures as function of time lag after the mid time of the sunspot cycle.}
\end{figure}

\section{Results}

\subsection{Correlations with time lag}

As a background for the investigation of possible relations between SCL and temperature in sunspot periods, we determined the correlation between SCL and temperature for variable lags of an 11-year time window.
 We calculated 11 year running mean temperatures for the selected data sets, and correlated this with the SCL with lags from zero to 13 years, calculated from the middle time for each solar cycle.
 The starting point could also have been chosen as the year of solar maximum or the end year of the sunspot cycle. However, we selected the middle time, since this gave the possibility to check correlations with the same sunspot cycle (lag=0).

 A lag is defined as later in time (in the future). As input temperature files we used the temperature series shown in figures 4-19, lower panels.

The result is displayed in figure 3. The North Atlantic locations (panel A) show a systematic better negative correlation up to a maximum negative for lags 9-12 years. 
For the 3 locations in southern Norway (panel B), the tendency is the  same, with a maximum negative correlation between 9 and 11 years. 
For Northern Norway and Archangel, the situation is somewhat different: The relations are not as steep as for the North Atlantic stations.
Three of the stations (Bod\o , Troms\o , and Karasjok) shows minima with 9-10 years lags, but a station further east, (Vard\o , 31E)  shows almost equal minima after 3 and $ \backsim $10 years lags. 
Finally Archangel  (41E), shows only one minimum after 2-3 years. 
We interpret this as the North Atlantic climate is dominating in Norway, except for the easternmost station (Vard\o ) where another climate zone  (boreal) appears. This type of climate dominates for Archangel further east. 
This means that if we calculate temperature averages for stations from one geographical area, we may mix climate zones, and for hemispherical or global temperatures, we may have a mixture of different time lags and the maximum correlation may therefore not be well defined.

Correlations with time lags for area-averaged temperatures are displayed in panel D, where results for Norway, Europe60 and HadCRUT3N series are shown. 
For Norway, the maximum negative correlation is for 11 years time lag and well defined, while the Europe60 average shows two minima of $ \backsim $2 and 10 years time lags, signaling the inclusion of eastern (boreal) climate stations. 
For this data set the highest correlation is after 2 years, which is mostly inside the same solar cycle.
The HadCRUT3N series shows only one shallow minimum at 6 - 12 years time lag -- apparently a mix of many time lags.


\subsection{Temperature trends}

The temperature series are analyzed with least square fits to the linear relation $ y=\beta x + \alpha $, where $ y $ is temperature or temperature anomaly (for Europe60 and HadCRUT3N), and $ x $ is time or length of sunspot cycles (SCL or PSCL). Results for data sets analyzed are shown in table 1.

\begin{table}
Table 1. Temperature trends and predictions for solar cycle 24

\begin{tiny}

\begin{tabular}{lcccccccccc}
\\
\hline
Place&Location& Start& $ \beta_{1} $ & $ \beta_{SCL} $ & $ \beta_{PSCL} $ &r$ ^{2}_{PSCL} $& $ t _{23} $ & $ t_{24} $& Conf int.
\\
&&yr&100yr$ ^{-1} $&yr$ ^{-1} $&yr$ ^{-1} $&&$ ^{\circ} $C&$ ^{\circ} $C&95\% \\
\hline

Oks\o y&58N,8E&1870&  0.44$ \pm $0.17 & 0.05$ \pm $0.14 & -0.28$ \pm $0.15 &0.25& 8.4 & -\\
Utsira&59.5N,5E&1868&   0.75$ \pm $0.21 & 0.03$ \pm $0.13 & -0.37$ \pm $0.12 &0.45& 8.3 & 6.9 & 6.1:7.7\\
Domb\aa s&62.4N,10E&1865&0.95$ \pm $0.23 & -0.01$ \pm $0.16 & -0.45$ \pm $0.15 &0.55& 2.6 & 0.9 &-0.1:1.8\\
Bod\o &67.5N,15E&1868&   0.78$ \pm $0.20 &-0.07$ \pm $0.15 & -0.42$ \pm $0.11 &0.56& 5.5 & 3.9 & 3.2:4.7\\
Troms\o &69.7,19E&1868& 0.36$ \pm $0.18 &-0.06$ \pm $0.10 & -0.29$ \pm $0.08 &0.56& 3.3 & 2.3 & 1.8:2.8\\
Karasjok&69.5N,25.5E&1876& 0.59$ \pm $0.43 &-0.11$ \pm $0.21 & -0.45$ \pm $0.18 &0.43&-1.1 &-2.6 & -3.8:-1.4\\
Vard\o  &70.4N,31E&1856& 0.91$ \pm $0.18 &-0.16$ \pm $0.17 & -0.38$ \pm $0.12&0.44& 2.3 & 0.8 &-0.1:1.7\\
\\
Norway&&1900&  0.56$ \pm $0.36 &0.11$ \pm $0.15 & -0.30$ \pm $0.12 &0.42& 1.82 &0.72 &-0.05:1.49\\
\\
Armagh  &54.3N,6W&1867& 0.44$ \pm $0.25 & 0.09$ \pm $0.13 & -0.29$ \pm $0.13 &0.33&10.1 & 8.8 & 8.0:9.7\\
Archangel&64.5N,41E&1881&1.4 $ \pm $ 0.3 &-0.18$ \pm $0.24 & -0.51$ \pm $0.21 &0.38& 1.5 &-0.5 &-1.9:0.9\\
Torshavn&62N,7W&1890& 0.58$ \pm $0.24 & -0.02$ \pm $0.12 & -0.34$ \pm $0.08 &0.67& 7.0 & 6.9 & 5.5:6.5\\
Akureyri&65.4,18W&1882& 1.16$ \pm $0.44 &-0.06$ \pm $0.25 & -0.71$ \pm $0.14 &0.72& 4.2 & 2.3 & 1.3:3.3\\
Svalbard&78.2,15.5E&1914& 1.64 $ \pm $1.1  & 0.69$ \pm $0.41 & -1.09$ \pm $0.31 &0.63&-4.3 &-7.8 &-9.6:-6.0\\
Nuuk    &64.1,51W&1881& 0.7$ \pm $0.6 &-0.12$ \pm $0.30 & -0.65$ \pm $0.24 &0.41&-0.7 & -2.3 &-4.0:-0.6\\
\\
Europe60&&1856& 0.62$ \pm $0.18 &-0.14$ \pm $0.12 & -0.29$ \pm $0.10 &0.39& 0.97 &-0.20 &-0.49:0.10\\
HadCRUT3N&&1856&0.47$ \pm $0.09 &-0.02$ \pm $0.09 & -0.21$ \pm $0.06 &0.49& 0.49&-0.38&-0.84:0.09$ ^{*} $\\
\hline\\
\end{tabular}
\\
Estimated standard deviations are given. $ ^{*} $Uncertain because of significant correlations in residuals (see table 2)
\end{tiny}
\end{table}

 $ \beta_{1} $ is the secular temperature trend for temperatures calculated from the average temperature in sunspot cycles including cycle 23 which terminated in 2008. The starting year of the temperature series are given in column 3.
 
$ \beta_{SCL} $ is the trend in the relation between the length of a solar cycle and the temperature in the same cycle. For all temperature series investigated, no trend significantly different from zero on the $ 2\sigma $ level has been found for $ \beta_{SCL} $.

$ \beta_{PSCL} $ is the trend between the length of one solar cycle and the temperature in the next cycle. We name this the previous solar cycle length (PSCL) relation, since we investigate the use of the length of a solar cycle as predictor for the temperature in the next.

$ r^{2}_{PSCL}$ is the estimated coefficient of determination based on the PSCL model.

Usually the degrees of freedom applied in the regression analysis are equal to the number of observations minus the number of parameters used in the modeling. 
However, since we have investigated the correlation as function of time lag after the middle time of a solar cycle, and found that maximum correlation exist about one sunspot cycle later, the number of degrees of freedom is reduced by one.
 
Our suggested model is rather simple: \textit{The air temperature in a sunspot cycle is a linear function of the length of the previous sunspot cycle (PSCL) in short: the $ t(PSCL) $-model.}

Table 1 shows that for all 16 series examined, except for Oks\o y, $ \beta_{PSCL} $ is significant negative on the  95\% level on the condition that the regression model applied gives independent residuals.

Analysis has been performed on the residuals, which are the differences between the temperature observations and the fitted temperatures. 
To check possible autocorrelation in the residuals a Durbin-Watson (DW) test \citep{DW50,DW51a,DW51b} has been applied. The results of the DW test are shown in Table 2.
On the 5\% significance level this test showed no significant autocorrelations in the residuals for 10 of the 16 series. 
Exceptions were the HadCRUT3N series, where significant autocorrelations were identified, and the Oks\o y, Domb\aa s, Bod\o , Norway and Archangel series where either significant autocorrelations or no significant autocorrelations were identified (indifferent cases). 

\begin{table}
Table 2. Durbin-Watson test on autocorrelations in the model residuals\\
\\
\begin{tiny}
\begin{tabular}{lccclcc}
\hline
&&Degrees&0.05 Significance&level&DW test&No correlation\\
Series&DW&freedom&D(L)      & D(U) & result&in residuals\\
\hline 
\\ 
Oks\o y&1.27&10&0.971&1.331&D(L)$ < $DW $<$ D(U)&Indifferent\\
Utsira &1.38&10&0.971&1.331&DW $>$ D(U)&Significant\\
Domb\aa s&1.03&10&0.971&1.331&D(L) $<$ DW $<$ D(U)&Indifferent\\
Bod\o &1.14&10&0.971&1.331&D(L) $<$ DW $<$ D(U)&Indifferent\\  
Troms\o &1.49&10&0.971&1.331&DW $>$ D(U)&Significant\\
Karasjok&1.54&9&0.927&1.324&DW $>$ D(U)&Significant\\
Vard\o &1.37&11&1.010&1.34&DW $>$ D(U)&Significant\\
\\
Norway&1.205&7&0.824&1.320&D(L)$<$DW$<$D(U)&Indifferent\\
\\
Armagh&1.55&10&0.971&1.331&DW $>$ D(U)&Significant\\
Archangel&1.24&10&0.927&1.324&D(L)$<$DW$<$D(U)&Indifferent\\
Torshavn&1.74&8&0.879&1.320&DW $>$ D(U)&Significant\\
Akureyri&1.59&9&0.927&1.324&DW $>$ D(U)&Significant\\
Svalbard&2.15&6&0.763&1.332&DW $< 4 -$ D(U)&Significant\\
Nuuk&1.67&9&0.927&1.324&DW $>$ D(U)&Significant\\
\\
Europe60&1.48&11&1.01&1.340&DW $>$ D(U)&Significant\\
HadCRUT3N&0.86&11&1.01&1.340&DW $<$ D(L)&Not significant\\  
\hline  
\end{tabular}

The Durbin-Watson lower D(L) and upper bounds D(U) numbers are given in \citet{SW77}. 

The number of parameters $ k $, is 1 in all cases since the regression constant is excluded. 

The number of observations in the DW test, is in all cases equal the number of degrees of freedom plus 2.
\end{tiny}
\end{table}

\begin{figure}[htb]
\epsfysize=90mm 
\epsfbox{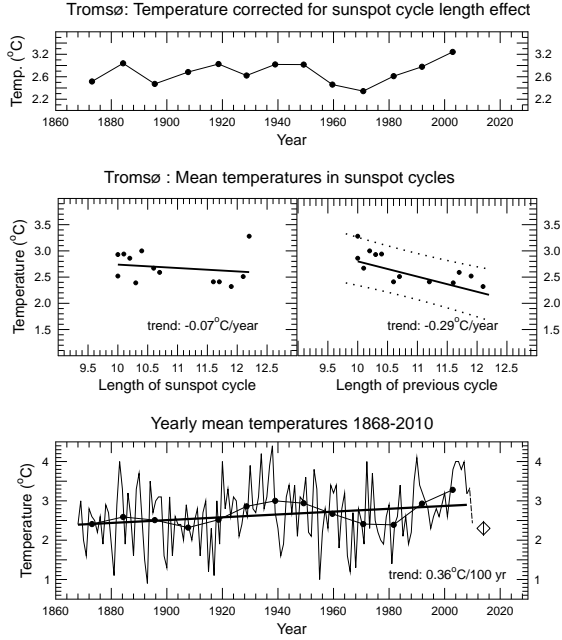}
\caption{Troms\o , Norway (Coastal city 70N): Lower panel: Annual mean temperatures and mean temperatures in sunspot cycles (dots) with linear trend. The broken line shows temperatures measured so far in SC24. 
The center two panels show temperatures in sunspot cycles versus SCL (left) or PSCL (right). In the right center panel the 95\% confidence interval is indicated with dotted lines.
The lower panel shows the forecasted value for SC24 (diamond) with standard deviation error bar. 
The top panel shows the residuals from the $ t(PSCL)$ model, and represents what may be left for other regressors and noise to explain.}
\end{figure}
For each of the 10 series with no autocorrelation in the residuals, the estimated trends are significant different from 0 since the absolute values of the estimated trends are greater than the estimated standard deviation, $\sigma$, multiplied by $t_{0.025}(f)$ - the 0.025 percentile in the Student distribution with $f$ degrees of freedom. 
Since the residuals are not autocorrelated, the degrees of freedom for the Student distribution are equal to the number of observations minus 3. As mentioned above  one additional degree of freedom is subtracted because of inspection of the data on beforehand.
Hence the degrees of freedom vary between 6 and 11, and the 0.025 percentile in the Student distribution varies between 2.20 and 2.45.

\begin{figure}[htb]
\epsfysize=90mm 
\epsfbox{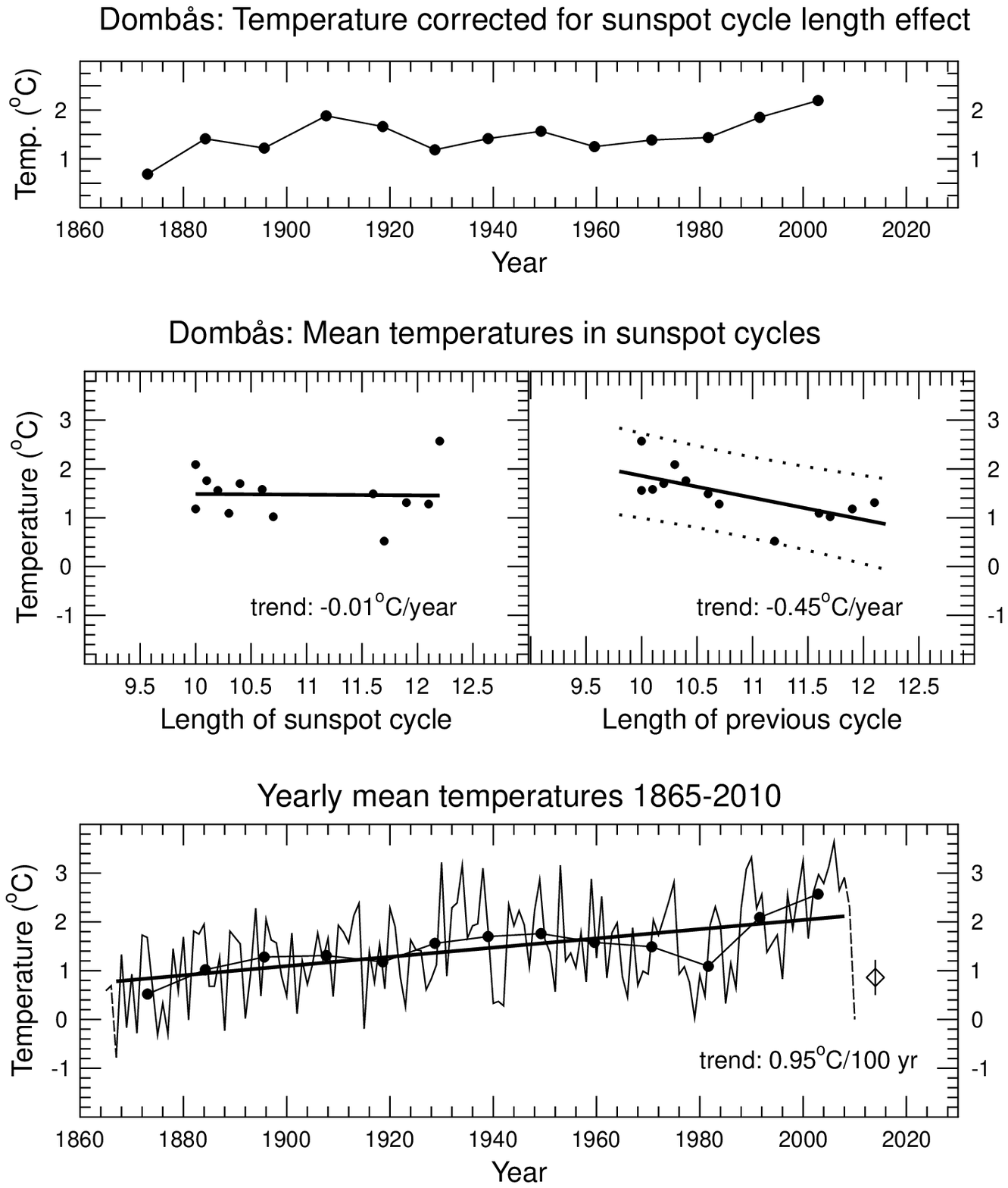}
\caption{Domb\aa s, Norway (Inland, mountain valley 62N), explanation as in figure 4}
\end{figure}

The number of average temperature observations for Bod\o\, and Domb\aa s is 13. Then $t_{0.025}(10)=2.23$. 
Table 1 shows that the absolute value of the trend estimate for Bod\o\, is $\sim 3.8 \sigma $ and for Domb\aa s $ \sim 3\sigma$. 
The 0.025 percentile for the Student distribution with 3 degrees of freedom, $t_{0.025}(3)=3.18$. 
It is no reason to believe that possible, but not significant autocorrelations in the residuals for Bod\o\ or Domb\aa s will reduce the degrees of freedom from 10 to 3 or less. Hence, the temperature trends for Bod\o\ and Domb\aa s are assumed to be significantly different from 0. 
For Archangel we find $ \beta_{PSCL}/\sigma_{\beta_{PSCL}} = 2.43 > t_{0.025} (10)=2.23$. Hence, also the Archangel temperature model is assumed to give significant results on a 95 \% significance level.

Analysis of the Oks\o y series does not give satisfactory results. Hence, because of a possible dependency in the residuals and a significance level lower than for the other series, it is not possible to estimate a 95 \% confidence level for the forecasts in this case. 

For the average temperature of Norway we get $ \beta_{PSCL}/\sigma_{\beta_{PSCL}} = 2.50 > t_{0.025} (7)=2.377$, which indicates that this is a significant result on the 95\% significance level.

\begin{figure}[htb]
\epsfysize=90mm 
\epsfbox{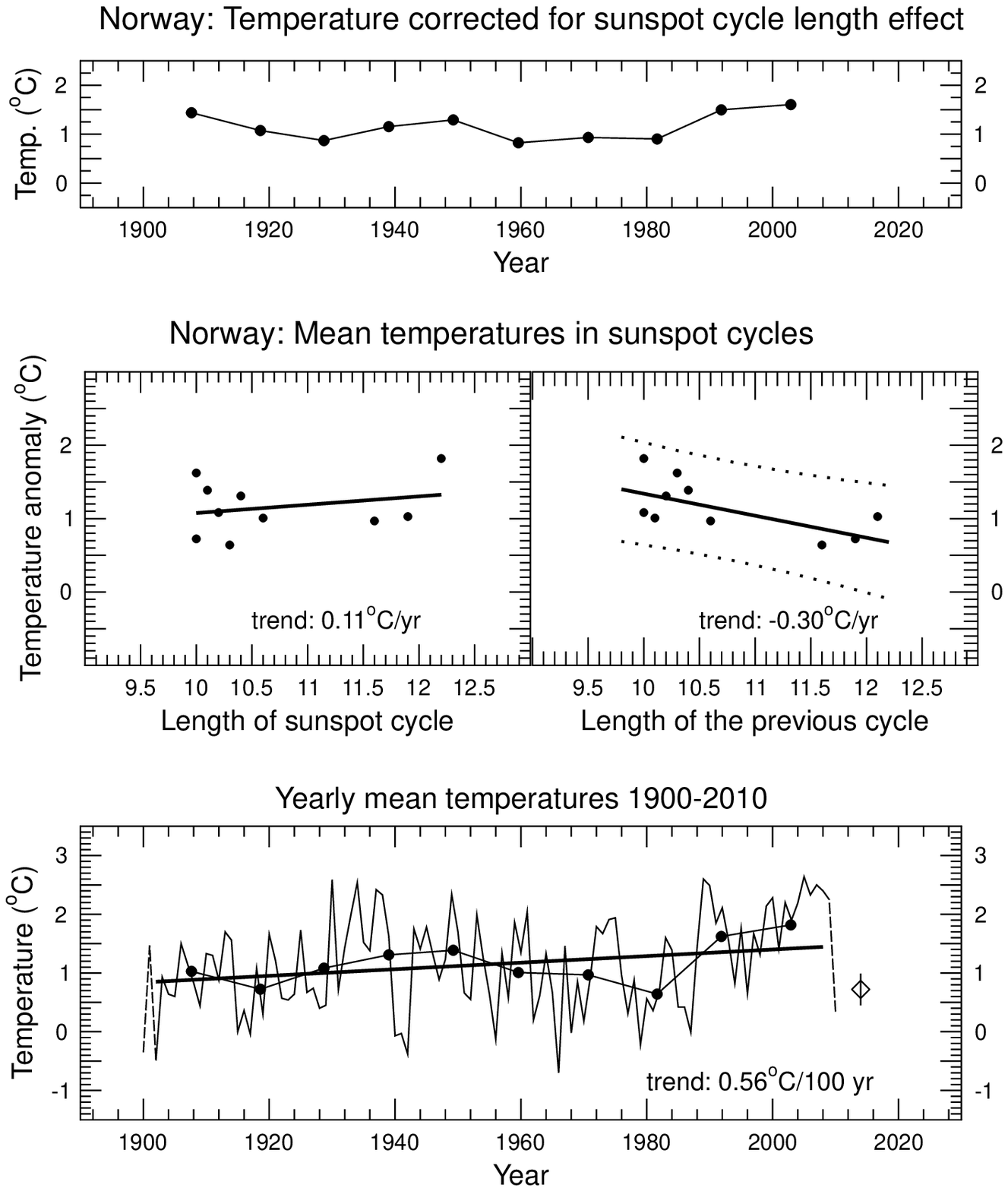}
\caption{Norway, average yearly temperatures, explanation as in figure 4.}
\end{figure}

The absolute value of the estimated HadCRUT3N temperature trend is $\sim 3.5 \sigma $. 
The number of temperature observations is 14 and the number of degrees of freedom 11. Then $t_{0.025}(11)=2.20$. 
Even with a reduction of number of freedoms from 11 to 3, the temperature trend will be significant different from 0. 
However, in this case there is reason to be more careful since significant autocorrelations in the residuals are identified.

The regression analysis indicates clearly that the temperatures in the PSCL model have significant trends different from 0 for 14 out of the 16 data series. 
It is also reasons to believe that the trend is significantly different from 0 for the HadCRUT3N temperature, but there is some uncertainty because of significant autocorrelations in the residuals.

A 95\% confidence interval of the estimated trend is given by respectively subtracting and adding the factor $\sigma\times t_{0.025}(f)$ to the estimated trend. 
This is the 95\% confidence interval for the temperature forecasts shown in the last column in table 1. 
For Oks\o y no confidence interval is estimated.
The next 14 confidence intervals are considered to be precise, while the last one for HadCRUT3N can be considered to be an approximate 95\% confidence interval.
We conclude that 15 of our 16 series support the PSCL regression relation.

\begin{figure}[htb]
\epsfysize=90mm 
\epsfbox{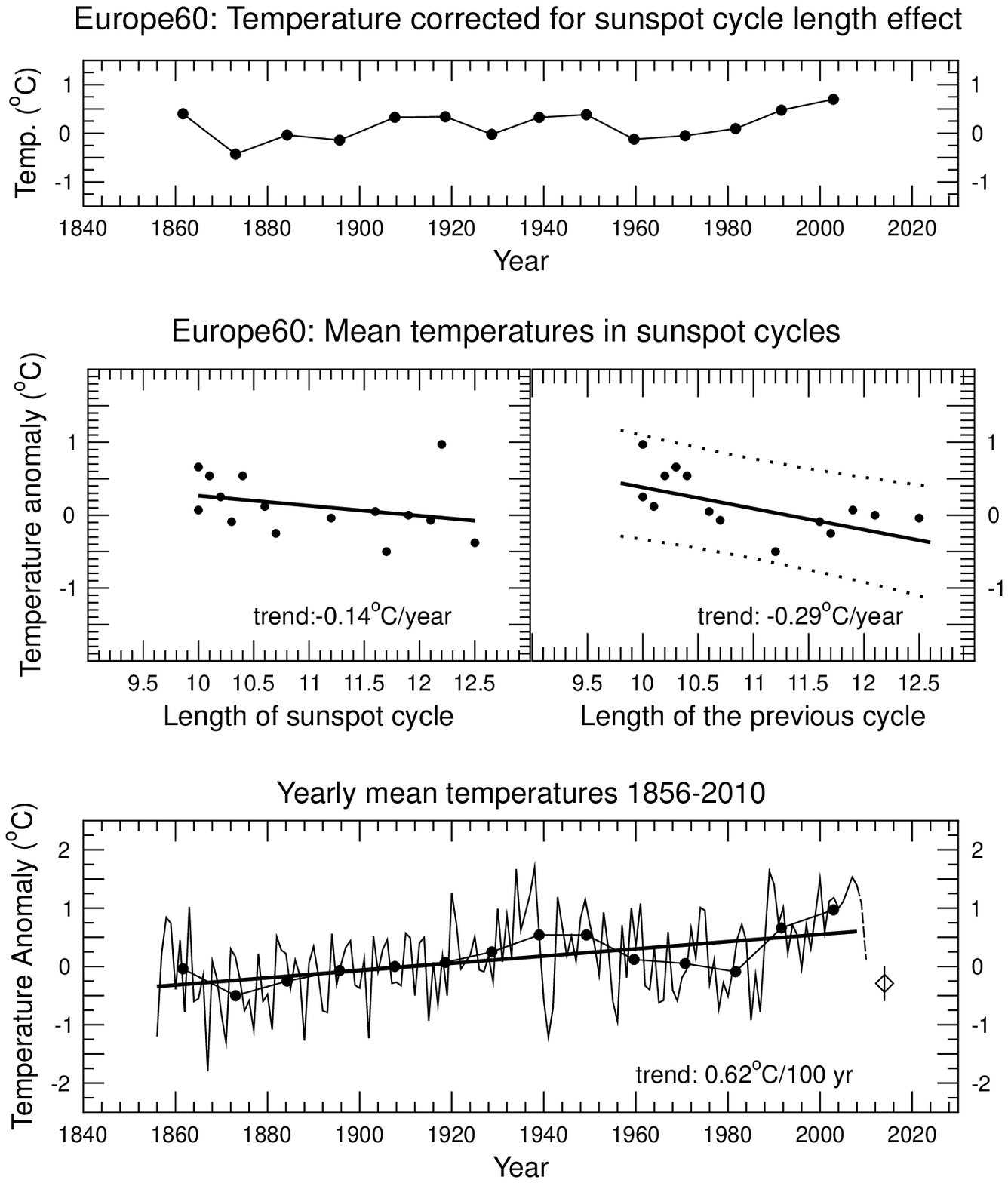}
\caption{Temperature anomaly for 60 stations on land in Europe, mostly located outside large cities. A list of the stations is given in the Appendix, explanation as in figure 4.}
\end{figure}

The coefficient of determination ($ r^2 $) indicates how much of the variance can be attributed to the regressor. Our results show that it varies from 0.25 at Oks\o y, at the southern tip of Norway, to 0.72 at Akureyri at the bottom of a large fjord in Northern Iceland. The highest values are found on islands and in northwestern coastal regions of the North Atlantic. Lower values are found inland, and distant from the main Atlantic Ocean currents. 

\subsection {Local temperature series} 
The results for two stations in Norway, one coastal north (Troms\o ) and one inland south (Domb\aa s), are shown in figures 4 and 5. 
For these two locations, and the others investigated and shown as figures 8-18 the $ \beta_{SCL} $-values are not significantly different from zero. 
This is demonstrated in the center left panels in the figures, which show the relations between the temperature and length of the same sunspot cycle. 
On the other hand, significant relations are found between PSCL and temperatures, which are shown in the center right panels.  
The residuals from the PSCL relations  (sorted time wise)  shown in the top panels, indicate the variability that remains to be explained either by other regressors or noise.

\begin{figure}[htb]
\epsfysize=90mm 
\epsfbox{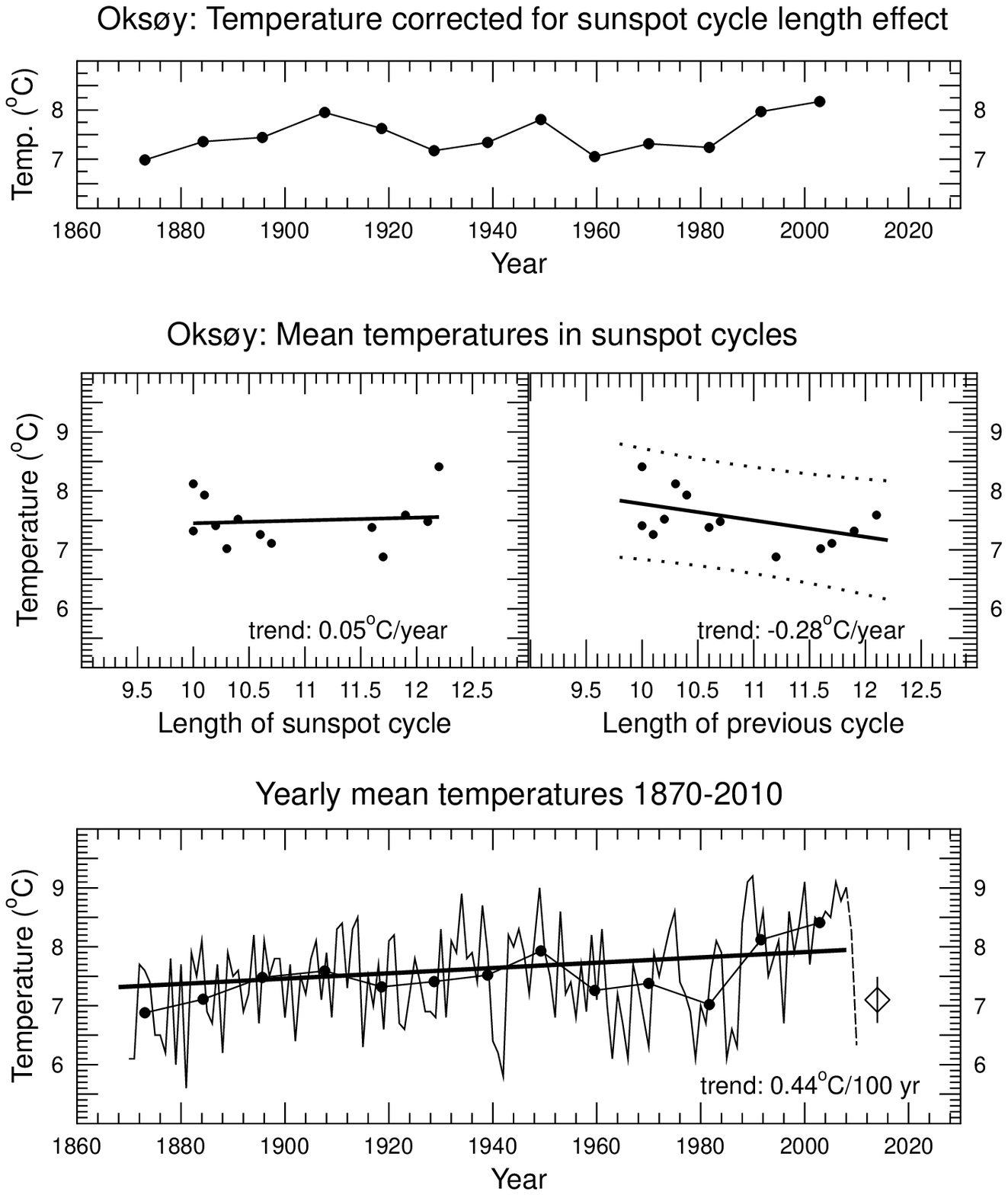}
\caption{Oks\o y lighthouse, Norway, average yearly temperatures, explanation as in figure 4. The confidence interval indicated in the center right panel is probably less than 95\% in this case.}
\end{figure}

For Troms\o \, (figure 4) the secular temperature trend before application of the PSCL-model is reduced from $ \beta_{1} =0.0036^{\circ} $Cyr$^{-1}$ to $ \beta_{2} =0.00096\pm 0.0004^{\circ} $Cyr$^{-1}$, where $ \beta_{2} $ is the secular trend in the residuals after removal of the sunspot cycle effect (PSCL-model). The coefficient of determination $ r^{2}=0.56$ indicates that more than half of the temperature variations may be attributed to solar activity. 

For Domb\aa s, situated in a mountain valley in southern Norway, (figure 5) the coefficient of determination is $ r^{2}=0.46 $, indicating that slightly less than half the temperature variations are related to solar activity as modeled by PSCL.  For Utsira, which is a lighthouse on the western coast of Norway (figure 9) $ r^{2}=0.45 $. The similar results indicate that the Atlantic currents may influence both locations. On the other hand, for Oks\o y (see figure 8), a lighthouse on the southern tip of Norway, we found that the PSCL-relation was not significant on the 95\% level, and this is reflected in $ r^{2}=0.25 $, which is the smallest solar contribution for the locations investigated. 
This location is influenced by different ocean streams that may cancel out the effect.

\begin{figure}
\epsfysize=90mm 
\epsfbox{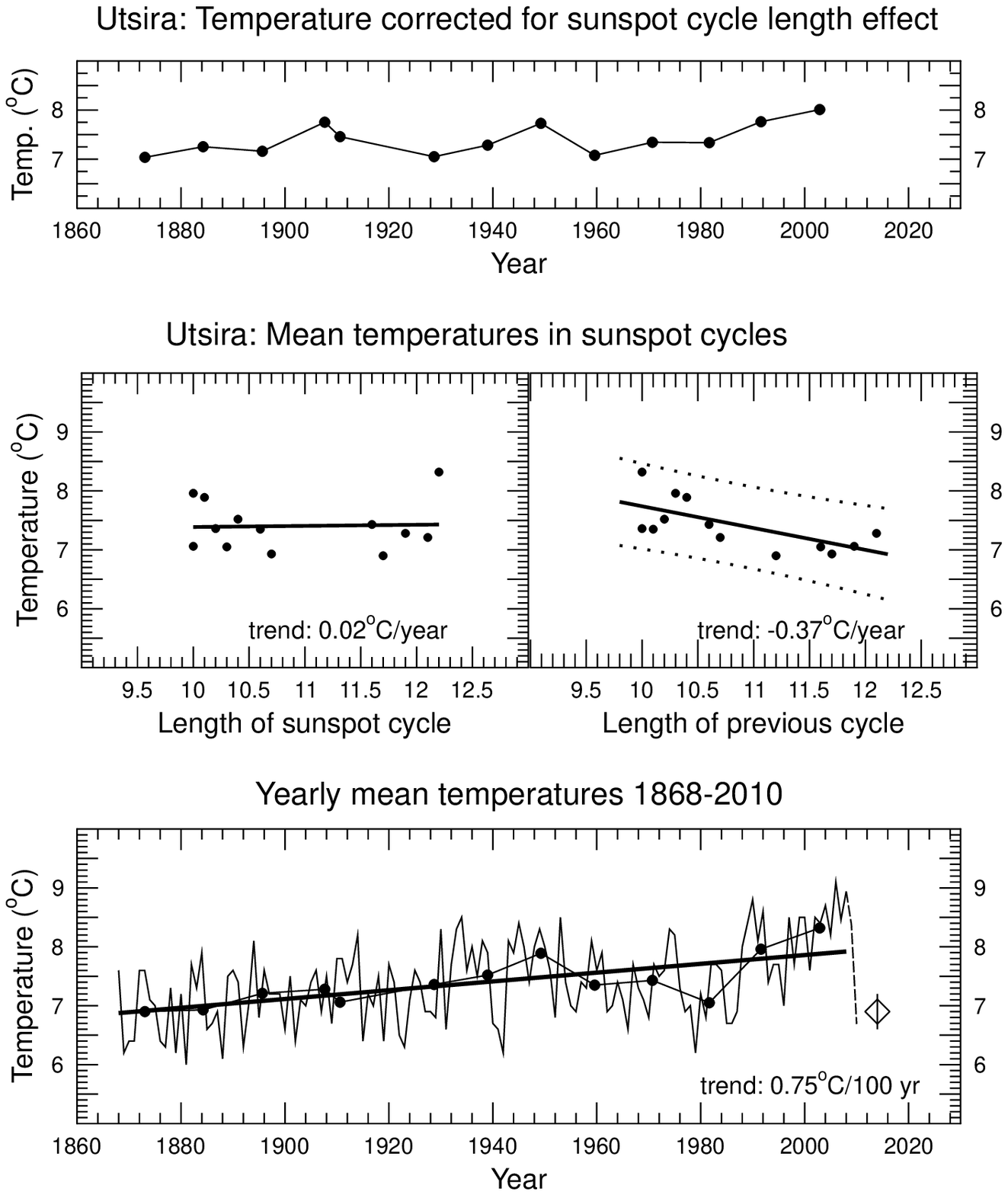}
\caption{Utsira lighthouse, Norway, average yearly temperatures, explanation as in figure 4.}
\end{figure}

In figures 4-19 are also shown (with diamonds) predicted average temperatures for SC24 based on the PSCL-model. 
Temperatures measured for 2009 and 2010 are shown as broken lines. For many locations the mean annual temperatures have decreased in the first years of SC24. 
In some of the figures yearly observations done before the first SC included in the analyses are connected with broken lines.

The average solar (PSCL) contribution to the variability for the 4 northern  Norwegian stations in table 1 is 50\%, compared to 42\% for the 3 stations located in southern Norway.  
Moving further west and north, we find that the stations at Torshavn, Akureyri and Svalbard (figures 15-17), have the maximum solar contribution with $ r^{2} = 0.63-0.72 $.

For Archangel (figure 13) we get $ \beta_{PSCL} = -0.51 \pm 0.21$, which is in line with the result of  \citet{AR10} who got $ \beta  = -0.6$, based on PSCL calculated from both the time between maxima and between minima. Archangel displayed the highest correlation with a time lag of only 2 years (figure 3). This may indicate a boreal climate different from the Atlantic, because of the shorter reaction time to a solar pulse.

\begin{figure}
\epsfysize=90mm 
\epsfbox{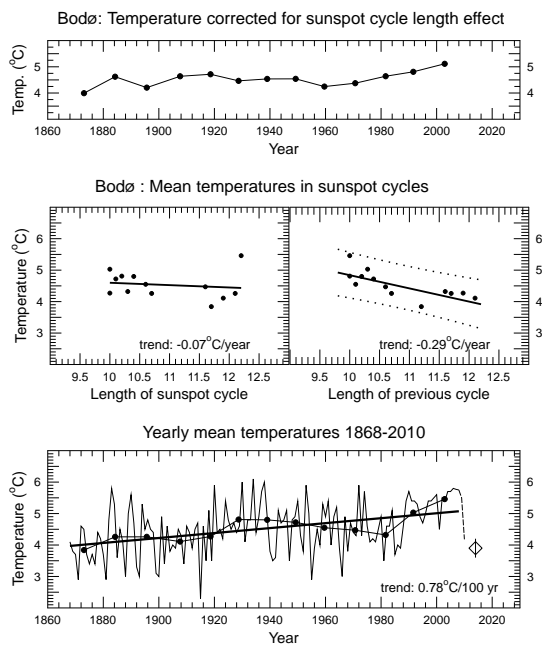}
\caption{Bod\o , Norway, average yearly temperatures, explanation as in figure 4.}
\end{figure}

For Armagh (figure 14) we get  $ \beta_{PSCL} = -0.29 \pm 0.13$, which is smaller than the value $ \beta = -0.5$ determined by \citet{BJ94}. 
Their larger value may come from their smaller range in SCL values due to the 1-2-2-2-1 smoothing applied by them.

The secular trend is determined with least significance for Svalbard (78N), which also has the largest trend, due to Arctic amplification. The Svalbard temperature record is based on measurements at the coast of the large fjord Isfjorden, and the temperature is therefore affected by both regional and local sea ice conditions. 
It is somewhat surprising that the coefficient of determination $ r^{2} =0.63$, which indicates that solar activity plays an important role for this Arctic location far North of the Polar Circle.
The series for Svalbard is also much shorter than the other series, but is still the longest in the high Arctic.
The temperature measured at Svalbard has already shown sign of decline as predicted. 
 
\begin{figure}
\epsfysize=90mm 
\epsfbox{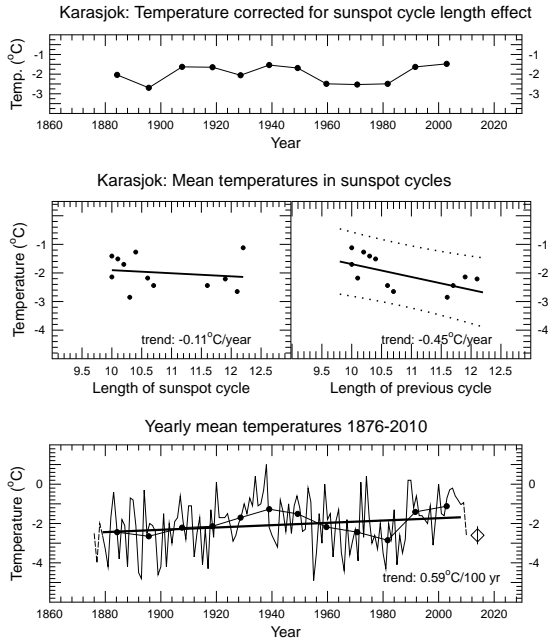}
\caption{Karasjok, Norway, average yearly temperatures, explanation as in figure 4.}
\end{figure}

A more detailed analysis of the Svalbard temperature series has been performed by \citet{SSH12}, who repeated this analysis with weighted averages and also analyzed each season separately. 
They confirmed the values of the correlation coefficient $ r $ with a bootstrap analysis, and got similar results as reported here for the coefficient of determination $r^{2}$. 
Only the yearly average and the winter season (DJF) temperatures showed no autocorrelations in the residuals. 
The winter temperatures gave $ r^{2} = 0.67 $, which indicates that the solar influence takes place also when the Sun is completely under the horizon.
Indirect lagged contributions by means of the ocean current and warm air advected from the south are the only possible explanations for the correlations determined. 

 For Nuuk on Greenland (figure 18), the secular trend ($ \beta_{1} $) is not significant, and disappears almost completely ($ \beta_{2} =0.0006\pm 0.0002^{\circ} $Cyr$^{-1}$) when corrected  for solar activity. 
The Nuuk station is located near the coast in West Greenland, and is affected by regional sea surface temperatures. 
 The sea between the Labrador island and West Greenland has shown a marked warm anomaly for most of 2010.
 This has resulted in a temperature increase the last two years (figure 18), opposite the prediction in table 1. 

\begin{figure}
\epsfysize=90mm 
\epsfbox{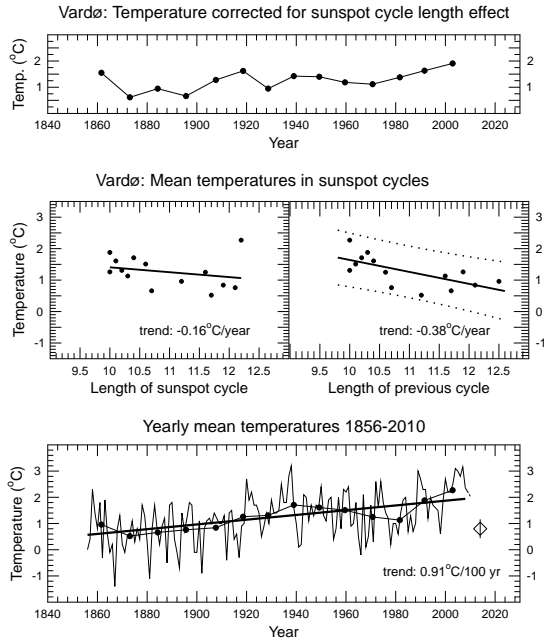}
\caption{Vard\o , Norway, average yearly temperatures, explanation as in figure 4.}
\end{figure}

\begin{figure}
\epsfysize=90mm 
\epsfbox{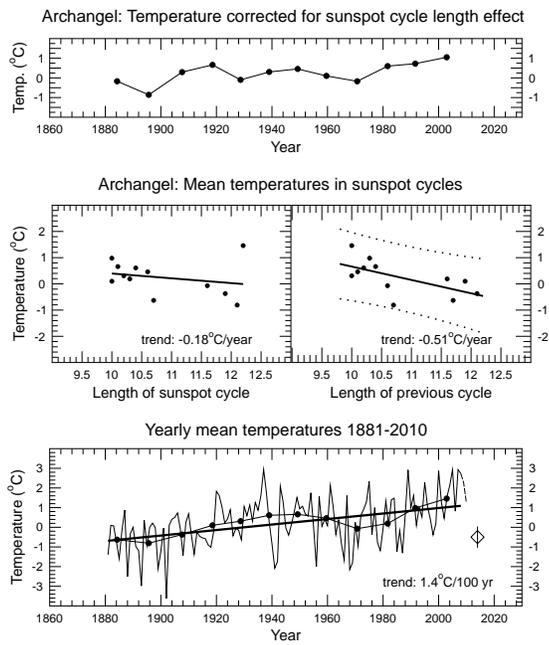}
\caption{Archangel, Russia, average yearly temperatures, explanation as in figure 4.}
\end{figure}

\begin{figure}
\epsfysize=90mm 
\epsfbox{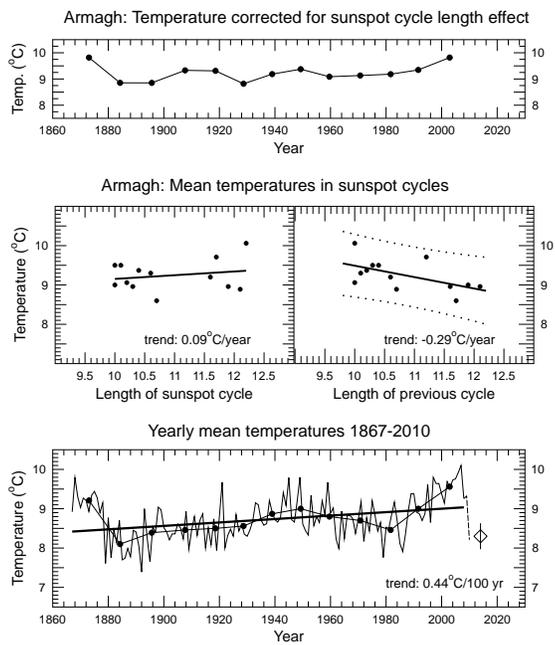}
\caption{Armagh, Northern Ireland, average yearly temperatures, explanation as in figure 4.}
\end{figure}

\begin{figure}
\epsfysize=90mm 
\epsfbox{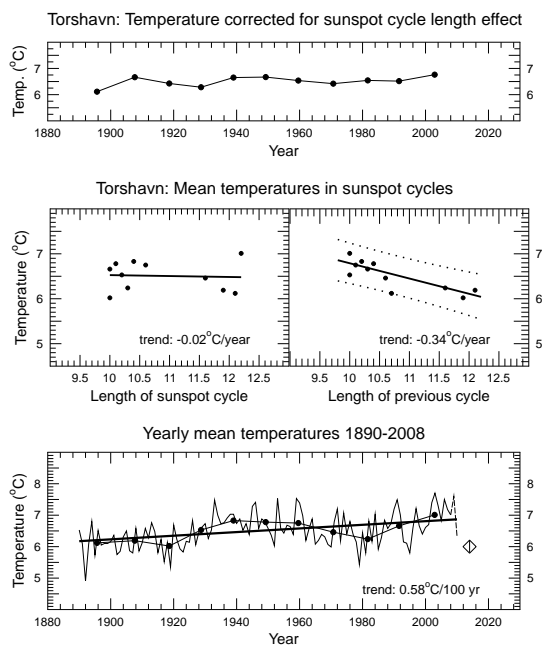}
\caption{Torshavn, Faroe Islands, average yearly temperatures, explanation as in figure 4.}
\end{figure}

\begin{figure}
\epsfysize=90mm 
\epsfbox{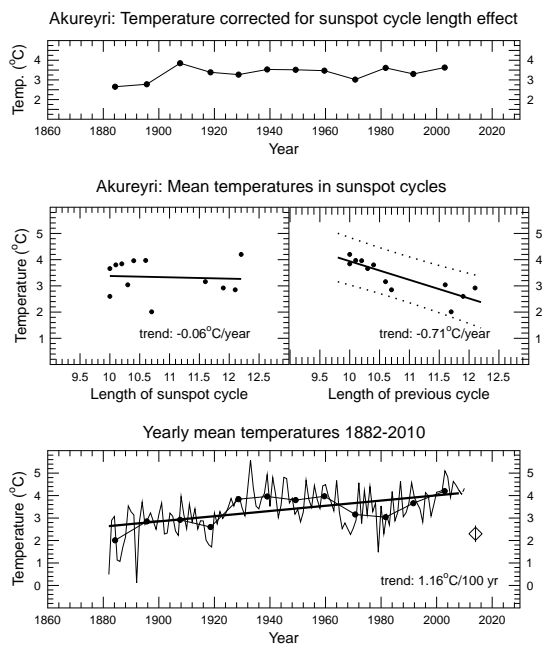}
\caption{Akureyri, Iceland, average yearly temperatures, explanation as in figure 4.}
\end{figure}

\begin{figure}
\epsfysize=90mm 
\epsfbox{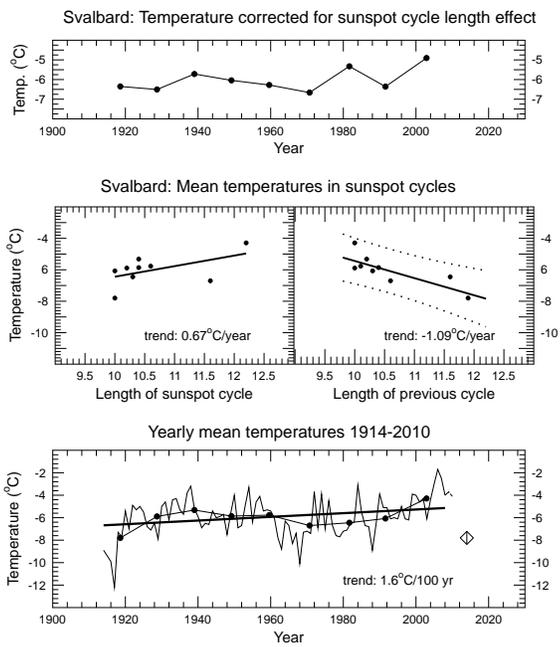}
\caption{Svalbard (Longyearbyen), average yearly temperatures, explanation as in figure 4.}
\end{figure}

\begin{figure}
\epsfysize=90mm 
\epsfbox{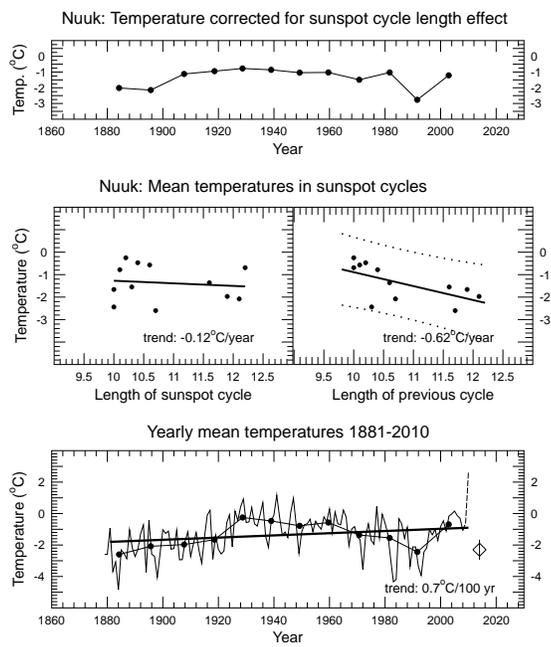}
\caption{Nuuk, Greenland,  average yearly temperatures, explanation as in figure 4.}
\end{figure}

\subsection{Area averaged air temperature series}

\begin{figure}[htb]
\epsfysize=90mm 
\epsfbox{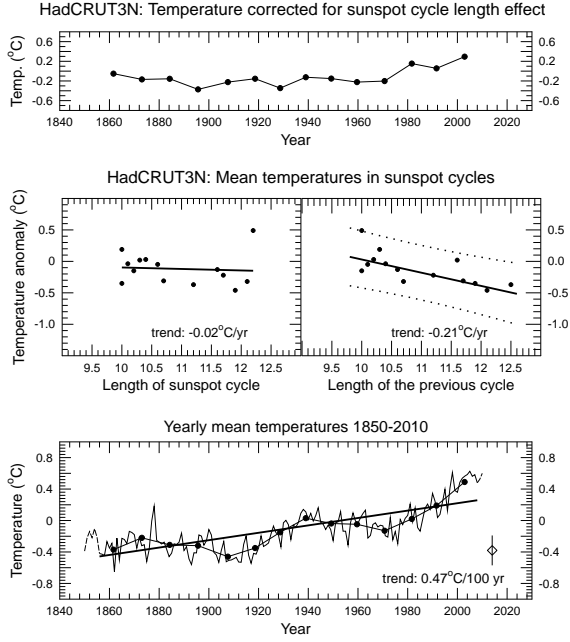}
\caption{HadCRUT3N, average yearly temperature anomalies, explanation as in figure 4.}
\end{figure}

The yearly average temperature for Norway calculated from 1900  onwards, is shown in the bottom panel of figure 6. A significant relation with $ \beta_{PSCL} = -0.30\pm 0.12 $  reduces the secular trend with a large fraction to $ \beta_{2} =0.0020\pm 0.0002^{\circ} $Cyr$^{-1}$. Solar activity may contribute at least 42\%.

For the average temperature for 60 European stations, mostly located outside big cities, we find a significant relation with $ \beta_{PSCL} = -0.29 \pm 0.10 $ (figure 7) and $ r^{2} =0.39$. HadCRUT3N temperature anomalies gives $ \beta_{PSCL} = -0.21 \pm 0.06 $, and  $ r^{2}=0.49 $, but in this case also other regressors may contribute significantly.

As seen in figures 6 and 7, the Norwegian and Europe60 average temperatures have already started to decline towards the predicted SC24 values, while the HadCRUT3N temperature anomaly has shown no such decline yet (figure 19).

\subsection{Testing the PSCL-model forecast}
To check the robustness of the forecast with the PSCL-model we have for all 16 datasets removed the last entry, which is the SC23 temperature, and used the remaining observations for generating a forecast of t(SC23). 
The result is that the forecast for 11 of the 15 series with 95\% confidence intervals are within this interval. For 3 of the series (Domb\aa s, Svalbard and HadCRUT3N, the forecasted temperature is 0.1$ ^{\circ} $C above the forecasted range, and in the case of Bod\o  , the deviation is 0.2$ ^{\circ} $C. 
If the SC23 had been a fraction shorter, all observed values would be inside the 95\% range.
It should be remarked that SC22 for a long time was listed in \citet{ng11} with a length of 9.6 or 9.7 yr, which was used in other analysis \citep{AR08,Th09}. According to earlier lists \citep{Th09} May 1996, which is the mathematical minimum for SC93, was in the beginning used, while a consensus among solar scientists moved the minimum to September 1996, which made SC22 longer (10.0 yr) and SC23 shorter (12.2 yr). This difference in definition of SCL from the original by \citet{Wa39} may explain the observations outside the confidence range. 

\section{Discussion}

The correlations found between the average temperature in a solar cycle and the length of the previous cycle, indicates a possible relation between solar activity and surface air temperature for the locations and areas investigated. 

The weak or non significant, correlations between the temperature and length of the same cycle, and the much stronger correlations for time lags of the order 10-12 years, makes the length of a solar cycle a good predictor for the average temperature in the next cycle.
 
The smoothing filters 1-2-2-2-1 and 1-2-1 or a length jump (sudden increase of the SCL), may be the reason that the previous good correlations between solar cycle length and Northern hemisphere land-temperature disappeared after 2000 \citep{TL00,Th09}.

We expect that the optimum delay is not exactly one solar cycle at some locations on the Earth, presumably due to oceanographic effects. 
This is confirmed by our determination of the correlations between SCL and temperature lags which appears to be at a maximum negative value for 10-12 year lags around the North Atlantic, but shorter at eastern locations as Vard\o\, on the Barents sea and Archangel in Northern Russia (figure 3).

Variations in temperature lags may not give precisely one solar cycle lag as the optimum relations between area averaged temperatures and SCL, but both for the Europe60 station average and HadCRUT N acceptable correlations are found with PSCL - and reasonable good  predictions are possible.

The PSCL-model predicts a temperature drop of 1.3 to 1.7$ ^{\circ}$C for single Norwegian stations analyzed from SC23 to SC24. For the average Norwegian and Europa60 temperatures the  temperature drop from SC23 to SC24 is 1.1-1.2 $ ^{\circ}$C. For HadCRUT3N the predicted temperature drop is 0.9$ ^{\circ}$C. 95\% confidence intervals for the predicted temperatures in SC24 are given in table 1, last column.
The Arctic cooling as predicted here may be converted into a global cooling, which is a factor 2-3 lower due to the Artic amplification of temperature differences \citep{MB02}. Tnis means a global cooling of the order 0.3-0.5$  ^{\circ}$C.  We may also expect a more direct cooling near Equator due to the response to reduced TSI with the weaker solar cycles in the near future \citep{Pe07,Me09,RR09}.

The PSCL relation is determined for the period 1850-2008 when the PSCL on average has shortened from cycle to cycle in relative small steps. and the Earth has warmed. The large increase in SCL from SC22 to SC23 signals a temperature drop, which may not come as fast as predicted because of the thermal inertia of the oceans. 
The warming has taken place over 150 years - cooling of the same order may require some decades to be realized. 
 
The temperatures in the North Atlantic and on adjacent land areas are controlled by the Atlantic Multidecadal Oscillation (AMO), which has in the instrumental period from 1856 exhibited a 65-–80 yr cycle (0.4$ ^{\circ} $C range), with warm phases at roughly 1860–-1880 and 1930–-1960 and cool phases during 1905-–1925 and 1970-–1990 \citep{KR00,GG04}. 
This period is proposed related to a 74 yr period, which is a sub harmonic of the 18.6 yrs lunar-nodal-tide  \citep{YT08}. 
The AMO may affect the temperatures at the locations we have analyzed, but a preliminary investigation by the authors have found considerable variations in dominant periods from place to place, which we interpret as due to local variations related to geography, dominant weather patterns, and distance from the ocean currents. Inclusion of periodic variations may improve the forecasts, and we plan to investigate this further.

The coefficient of determination $ r^{2} $ is a measure of the relative contribution of solar activity modeled by the PSCL-model, to the temperature variations. The range of  $ r^{2} $ is from 0.25 (Oks\o y) to 0.72 (Akureyri).
The highest values (Torshavn: 0.62, Akureyri: 0.72 and Svalbard: 0.63) are found in or close to the North Atlantic ocean currents. This points to ocean currents as the mechanism of transport of the heat generated at southern locations by solar radiation. 

Even if the correlation is rather high between the actual temperature and the fitted temperature by use of regression modeling, the question arises of possibilities to improve the models.
Other regressors, which may be included, are atmospheric and  oceanic dynamics unrelated to the Sun or with different time lags, volcanic activity, GHG forcing, and anthropogenic activity of various kinds. 
The contribution from other regressors should be determined for each location, and may be quite different from area-averaged regressors. This is a question left for future investigations. 
However, there are clear limitations for including additional explanatory variables simply because there are limited numbers of observations available. 
Additional variables will improve the model fitting, but reduce the degree of freedom. 
 
This analysis shows significant dependency between the previous sunspot cycle length and the temperature. The established model is able to make significant forecasts with 95\% confidence limits for the present sunspot cycle.
There are reasons to believe that these results could be fundamental in further development of long-term forecasting models for the temperature. 

Looking at figure 1, which shows the variation in the length of solar cycles, we realize that short cycles like the one that ended in 1996, have only been observed 3 times in 300 years.
After the shortest cycles, sudden changes too  much longer cycles have always taken place, and thereafter there is a slow shortening of the next cycles, which take many cycles to reach a new minimum.
This recurrent pattern tells us that we can expect several long cycles in the next decades. 
Analysis of the SCL back to 1600 has shown a periodic behavior with period 188 yr, now entering a phase with increasing SCL the next $ \sim $75 yr \citep{RR09}.

 \citet{DJD11} concludes that the solar activity is presently going through a brief transition period (2000-2014), which will be followed by a Grand Minimum of the Maunder type, most probably starting in the twenties of the present century. 
Another prediction, based on reduced solar irradiance due to reduced solar radius, is a series of lower solar activity cycles leading to a Maunder like minimum starting around 2040 \citep{Ab07}.

A physical explanation for the correlations between solar activity parameters as the SCL and the temperatures on the Earth, and a possible decoupling of this relation the last few years has been investigated by \citet{SK03}.  
Their conclusion is that even if the proxies and direct measurements are scaled such that statistically the solar contribution to climate is as large as possible in the period 1856-1970, the Sun cannot have contributed to more than 30\% of the global temperature increase taken place since then, irrespective of which of the three considered channels is the dominant one determining Sun-climate interactions: 
Tropospheric heating caused by changes in total solar irradiance, stratospheric chemistry influenced by changes in the solar UV spectrum, or cloud coverage affected by the cosmic ray flux.

From correlation studies of 7 (not all global) temperature series for the period 1610-1970 \citet{DJDG10} found a solar contribution of 41\% to the secular temperature increase. Our results  are somewhat higher for Northern Hemisphere locations in the period 1850-2008.

Analyzing global temperature curves for periodic oscillations \citet{Sc10} concludes that the climate is forced by astronomical oscillations related to the Sun, and at least 60\% of the warming since 1970 can be related to astronomical oscillations. 
Looking at our figures 4-19, we can see sign of quasi-periodic variations in the lower panels, which have more or less disappeared in the upper panels. 
We may therefore suggest that SCL in some way is related to astronomical forcing.
 
Satellite observations by the Spectral Imager Monitor (SIM) indicate that variations in solar ultraviolet radiation may be larger than previously thought, and in particular, much lower during the recent long solar minimum. 
Based on these observations \citet{IS11} have driven an ocean-climate model with UV irradiance. 
They demonstrate the existence of a solar climate signal that affects the NAO (North Atlantic Oscillation) and produced the three last cold winters in Northern Europe and in the United States.

\section{Conclusions}

Significant linear relations are found between the average air temperature in a solar cycle and the length of the previous solar cycle (PSCL) for 12 out of 13 meteorological stations in Norway and in the North Atlantic. For 9 of these stations no autocorrelation on the 5\% significance level was found in the residuals. 
For 4 stations the autocorrelation test was undetermined, but the significance of the PSCL relations allowed for 95\% confidence level in forecasting for 3 of these stations.
Significant relations are also found for temperatures averaged for Norway, 60 European stations temperature anomaly, and for the HadCRUT3N temperature anomaly. Temperatures for Norway and the average of 60 European stations showed indifferent or no autocorrelations in the residuals. The HadCRUT3N series showed significant autocorrelations in the residuals.

For the average temperatures of Norway and the 60 European stations, the solar contribution to the temperature variations in the period investigated is of the order 40\%. 
An even higher contribution (63-72\%) is found for stations at Faroe Islands, Iceland and Svalbard. This is higher than the 7\% attributed to the Sun for the global temperature rise in AR4 \citep{AR4}. 
About 50\% of the HadCRUT3N temperature variations since 1850 may be attributed solar activity. However, this conclusion is more uncertain because of the strong autocorrelations found in the residuals. 

The significant linear relations indicate a connection between solar activity and temperature variations for the locations and areas investigated. 
A regression forecast model based on the relation between PSCL and the average air tempereaure is used to forecast the temperature in the newly started solar cycle 24. This forecast model benefits, as opposed to the majority of other regression models with explanatory variables, to use an explanatory variable - the solar cycle length - nearly without uncrtainty. Usually the explanatory variables have to be forecasted, which of cause induce significant additional forecasting uncertainties.

Our forecast indicates an annual average temperature drop of 0.9$ ^{\circ} $C in the Northern Hemisphere during solar cycle 24. For the measuring stations south of 75N, the temperature decline is of the order 1.0-1.8$ ^{\circ} $C and may already have already started. For Svalbard a temperature decline of 3.5$ ^{\circ} $C is forecasted in solar cycle 24 for the yearly average temperature. 
An even higher temperature drop is forecasted in the winter months \citep{SSH12}.

Artic amplification due to feedbacks because of changes in snow and ice cover has increased the temperature north of 70N a factor 3 more than below 60N \citep{MB02}. An Artic cooling may relate to a global cooling in the same way, resulting in a smaller global cooling, about 0.3-0.5 $ ^{\circ} $C in SC24.

Our study has concentrated on an effect with lag once solar cycle in order to make a model for prediction. 
Since solar forcing on climate is present on many timescales, we do not claim that our result gives a complete picture of the Sun's forcing on our planet's climate.
\\ 


 {\it Acknowledgments} 
 
 A special thank to D. Archibald for valuable comments, and to the various agencies for providing data for analysis: Hadley Centre, Norwegian Meteorological Institute (e-Klima portal), www.rimfrost.no, and the GISS data base for the European stations, using the criterion "after combining sources at the same location", in addition to \O. Nordli for providing the homogenized Domb\aa s series. Finally we thank two referees for valuable comments, which have helped us improve the work significantly in two referee-cycles.


\newpage

\appendix
\section{The Europe60 stations}
\begin*{
Table 3. Location of Meteorological stations included}
\begin{tiny}

\begin{tabular}{lcclcclcc}

Arhangel'Sk   &64.5 N& 40.7 E &Karesuando   &68.5 N &22.5 E &Stornoway   &58.2 N& 6.3 W\\
Bj\o rn\o ya   &74.5 N&  19.0 W&Kem'-Port   &65.0 N& 34.8 E &Stuttgart   &48.8 N& 9.2 E\\
Bod\o   &67.3 N&  14.4 E &Kremsmuenster   &48.0 N &14.1 E&Tafjord   &62.2 N&  7.4 E\\
Brest   &48.5 N& 4.4 W &K\o benhavn   &55.7 N&  12.6 E&Tiree   &56.5 N& 6.9 W\\
Brocken   &51.8 N& 10.6 E  &Lerwick   &60.1 N& 1.2 W&Torshavn&62.0 N&6.8 W\\
De Bilt   &52.1 N& 5.2 E&Lomnicky Stit   &49.2 N &20.2 E&Tot'Ma   &59.9 N& 42.8 E\\
Dublin Airport   &53.4 N& 6.2 W  &L'Viv   &49.8 N& 23.9 E  &Trier-Petrisb   &49.8 N& 6.7 E  \\               
Elat'Ma   &55.0 N& 41.8 E&Mont Aigoual   &44.1 N& 3.6 E& Troms\o   &69.5 N&  19.0 E\\
Elblag   &54.2 N& 19.4 E&Murmansk   &69.0 N& 33.0E&Uppsala   &59.9 N& 17.6 E\\
Fichtelberg   &50.4 N& 12.9 E &Navacerrada   &40.8 N& 4.0 W&Utsira  &59.3 N&  4.9 E\\
Goteborg-Save   &57.8 N& 11.9 E &Oktjabr'Skij   &51.6 N& 45.5 E&Valentia Obs.   &51.9 N& 10.2 W\\
Gridino   &65.9 N& 34.8 E&Onega   &63.9 N& 38.1 E&Vard\o  &70.4 N& 31.1 E\\
Haparanda   &65.8 N &24.1 E &Pecs   &46.0 N& 18.2 E &Vf. Omu   &45.5 N& 25.4 E\\
Helsinki-Seutula&60.3 N &25.0 E &Praha-Ruzyne   &50.1 N &14.2 E&Vilnius   &54.6 N& 25.1 E\\
Hohenpeissenb   &47.8 N& 11.0 E &Reboly   &63.8 N& 30.8 E  &Visby Airport  &57.7 N& 18.4 E\\ 
Jan Mayen&70.9 N&8.7W&R\o ros   &62.3 N  &11.2 E& Vologda   &59.3 N& 39.9 E\\
Jungfraujoch&46.5 N& 8.0 E  &Saentis   &47.2 N &9.3 E&Vytegra   &61.0 N& 36.5 E\\
Kandalaksa   &67.2 N& 32.4 E&Ship M   &66.0 N &2.0 E  &Wlodawa   &51.5 N& 23.5 E\\
Kanin Nos   &68.7 N& 43.3 E&Sodankyla   &67.4 N& 26.6 E &Zugspitze  &47.4 N &11.0 E\\   
Karasjok  &69.5 N& 25.50 E &Stockholm   &59.3 N& 18.1 E&Zurich  &47.4 N& 8.6 E\\

\end{tabular}

\end{tiny}
\end*{}

\begin{figure}[htb]
\epsfysize=60mm 
\epsfbox{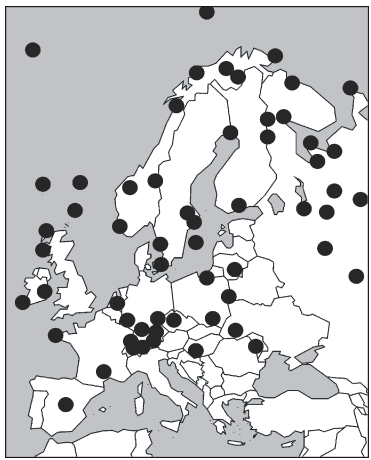}
\caption{The location of the  Europe60 stations}
\end{figure}

\begin{figure}
\epsfysize=50mm 
\epsfbox{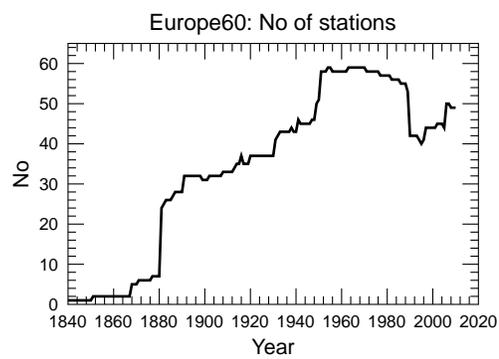}
\caption{The variation in numbers of stations in the Europe60 average}
\end{figure}


\end{document}